\documentclass[twocolumn,preprintnumbers,floatfix,prb,superscriptaddress]{revtex4}
\usepackage{graphicx}
\usepackage{graphicx}
\usepackage{dcolumn} 
\usepackage{bm}
\usepackage{epsfig}
\begin{document} 

\title{Electron Confinement, Orbital Ordering, and Orbital Moments\\
     in $d^0$-$d^1$ 
    Oxide Heterostructures}

\author{Victor Pardo}
 \email{victor.pardo@usc.es}
\affiliation{Department of Physics,
  University of California, Davis, CA 95616
}
\affiliation{
Departamento de F\'{\i}sica Aplicada, Universidade
de Santiago de Compostela, E-15782 Santiago de Compostela,
Spain
}

\author{Warren E. Pickett}
 \email{wepickett@ucdavis.edu}
\affiliation{Department of Physics,
  University of California, Davis, CA 95616
}


\begin{abstract}
The (SrTiO$_3$)$_m$/(SrVO$_3$)$_n$ $d^0-d^1$ multilayer system is studied with
first principles methods through the observed insulator-to-metal transition with
increasing
thickness of the SrVO$_3$ layer.  When correlation effects with reasonable magnitude
are included, crystal field splittings from the structural relaxations together
with spin-orbit coupling (SOC) determines the behavior of the electronic and
magnetic structures.  These confined slabs of SrVO$_3$ prefer $Q_{orb}$=($\pi,\pi$) orbital ordering
of $\ell_z = 0$ and $\ell_z = -1$ ($j_z=-\frac{1}{2}$) orbitals within the plane, accompanied 
by $Q_{spin}$=(0,0) spin
order (ferromagnetic alignment).  The result is a SOC-driven ferromagnetic Mott insulator.
The orbital moment of 0.75 $\mu_B$ strongly compensates the spin moment
on the $\ell_z = -1$ sublattice.
The insulator-metal transition for $n = 1 \rightarrow 5$ (occurring between $n$=4 and
$n$=5) is reproduced.
Unlike in the isoelectronic $d^0-d^1$ TiO$_2$/VO$_2$ (rutile structure) system and in spite
of some similarities in orbital ordering, no semi-Dirac
point [{\it Phys. Rev. Lett.} {\bf 102}, 166803 (2009)] is encountered, but the insulator-to-metal
transition occurs through a different type of unusual phase.
For n=5 this system is very near (or at) a unique semimetallic state in which the Fermi energy
is topologically determined and the Fermi surface consists of identical electron and hole
Fermi circles centered at $k$=0. The dispersion consists of what can be regarded as a continuum
of radially-directed Dirac points, forming a ``Dirac circle''. 

\end{abstract}

\maketitle

\section{Background}
Oxide heterostructures with a polar discontinuity across interfaces have 
attracted a great deal of attention recently, due to the unusual electronic 
behavior that can arise.\cite{nakagawa,pentcheva_jpcm} It is now becoming evident that heterostructures with
non-polar interfaces can also lead to unanticipated behavior, including 
low energy dispersion that is distinct from any previously known system.
The specific example is the $d^0$/$d^1$ interface system TiO$_2$/VO$_2$ 
that displays a point Fermi surface, from which semi-Dirac dispersion\cite{sD_prl} emerges.  Semi-Dirac  dispersion is characterized by
conventional, massive dispersion along one direction in the 
two-dimensional plane but massless
dispersion in the perpendicular direction. 

Although reminiscent of graphene,
the semi-Dirac system displays its own distinctive low energy 
properties.\cite{swapno_sD} The behavior is actually an electron confinement
phenomenon assisted by a particular orbital ordering, and these nanostructures 
display a peculiar metal-insulator transition\cite{vo2_tio2_mit} as the 
thickness of the d$^1$ oxide is increased. In these rutile structured oxides, 
the metal-insulator transition takes place through an intermediate semi-Dirac 
point when the thickness is approximately 
1 nm, where the system is neither insulating nor conducting and the Fermi 
surface is point-like.

Transition metal oxide perovskites with 3d$^1$ configuration are known to be
on the borderline between metallic and insulating, depending on the relative
sizes of several electronic energy scales, including the ratio between the 
on-site Coulomb repulsion U and the bandwidth W of the $d$ electrons, and the 
competition between magnetic energies and Jahn-Teller splittings. For small 
U/W, the material will be metallic, like the correlated metal\cite{srvo_goody} 
SrVO$_3$.  For a large value of the ratio U/W, the system 
will present a more localized behavior, and a Mott insulator will result, 
as in LaTiO$_3$ or YTiO$_3$.\cite{d1systems_04} Other energy scales may also 
affect, or even determine, the delicate balance.

Multilayers of SrVO$_3$ (SVO) and SrTiO$_3$ (STO)
have been grown on STO substrates by Kim {\it et al.}
and have displayed a transition\cite{svo_sto_ssc} from the typical insulating behavior of STO to
the metallic behavior of SVO as the number of layers of each constituent is increased.
Superlattices formed by films with 2 layers of STO and 6 layers of SVO [we will denote (SrTiO$_3$)$_m$/(SrVO$_3$)$_n$ (001) oriented multilayers as $m/n$]  
already show metallic behavior with a nearly flat resistivity curve with magnitude
close to that of SVO.
However, from the behavior of the resistivity, it was observed
that the 2/3, 2/4 and 2/6 films (increasing SVO thickness) show an insulator-metal
transition at temperatures
ranging from about 100 K for the 2/6 film to the approximately 230 K of the 2/4 and
2/3 films. 
It is to be noted also that only 2 layers of STO are not
enough to isolate the SVO slabs, as can be seen from the resistivity data for the 2/3 and 3/3
systems. In fact, the 2/3 system is already semi-metallic, but four 
layers of STO are enough to render interactions between neighboring SVO slabs negligible. 
Five SVO layers are needed to obtain a metallic state, as we show below from our calculations.

In this paper we extend our investigations of $d^0$-$d^1$ nanostructures by studying
this STO/SVO system,
focusing on the differences that the crystal structure, with its specific 
crystal field splittings, can cause.  We choose the most commonly studied
structure, perovskite, with the previously studied nanostructures with rutile
crystal structure.  We will compare different orbital orderings and magnetic arrangements,
and also study how the insulator to metal transition occurs with increasing
thickness of the $d^1$ material (SrVO$_3$).

\section{Computational methods}

Our electronic structure calculations were  performed within density functional 
theory \cite{dft} using the all-electron, full potential code {\sc wien2k} \cite{wien}  
based on the augmented plane wave plus local orbital (APW+lo) basis set \cite{sjo}.
The exchange-correlation 
potential utilized for the structure optimizations was the generalized gradient approximation (GGA) in the 
Perdew-Burke-Ernzerhof (PBE)\cite{gga}.  To deal with  strong correlation effects that are evident 
in SrVO$_3$ we apply the LSDA+U 
scheme \cite{sic1,sic2} including an on-site U and J (on-site Coulomb repulsion and exchange strengths) 
for the Ti and V $3d$ states.  The values U= 4.5 eV, J= 0.7 eV have been used for V to deal properly with 
correlations in this multilayered structure; these values are comparable to those used in literature 
for d$^1$ systems\cite{vo2_peierls2,tomczak,haverkort}. Our calculations 
show that a larger U, above 5.0 eV, give an incorrect insulating behavior of bulk SrVO$_3$ in 
cubic structure, hence overestimating electron-electron interactions.
Since Ti $d$ states never have any significant occupation, including U or not on Ti $3d$
orbitals has negligible consequence.
Spin-orbit coupling (SOC) effects on the valence and conduction states, which is discussed in the 
last section, have been introduced in a second 
variational procedure  built on the
scalar relativistic approximation.

\section{Treating correlation effects in bulk SrVO$_3$}
We have first performed calculations on bulk SrVO$_3$ (SVO) to establish the thick SrVO$_3$ limit of 
these nanostructures. Experimentally, SrVO$_3$ is a ferromagnetic (FM) metal\cite{srvo_magnetism} 
crystallizing in a cubic perovskite structure.\cite{srvo_struct} No distortion from cubic 
structure has been observed experimentally, which is consistent to its metallic character; a
Mott insulating $d^1$ system would be expected to distort due to orbital ordering. Our 
calculations show that the most stable structure based on GGA exchange-correlation (not including 
on-site Coulomb repulsion effects) is cubic with no distortions. However, a non-magnetic solution is 
obtained as a ground state within GGA. 

When correlations are introduced by means of the LDA+U method, the most stable structure is 
slightly distorted. This broken symmetry arises because the LDA+U method tends to promote integer 
occupations of one of the $t_{2g}$ orbitals (which becomes lower in energy, and preferentially occupied). 

Calculations were carried out with a tetragonal distortion in bulk SrVO$_3$ and with various values 
of the on-site Coulomb repulsion U. LDA+U calculations predict two possible orbital configurations: a 
FM configuration with all the t$_{2g}$ bands equally occupied and an AF solution with a substantial 
occupation of the d$_{xy}$ orbital. As U increases above 5 eV, the AF solution is more stable, even 
for undistorted cubic perovskite, leading to an incorrect AF insulating state. In all ranges of U 
studied (3 to 7 eV), a tetragonal distortion is more stable, whereas a simple GGA calculation leads 
to an undistorted cubic solution, in agreement with experiment. Even in the case of a FM solution, 
when a U is introduced in the calculations, a tetragonal distortion is obtained as a ground state, 
in disagreement with experimental observations. Using the correct structure obtained from both 
experiment and GGA calculations (which agree), the LDA+U method with values of U smaller than 5 eV 
leads to the correct FM metallic state as ground state.

Within GGA (U = 0) SVO is a cubic metal, with all the t$_{2g}$ orbitals are equally occupied. 
Within LDA+U, a tetragonal distortion causes the preferential occupation of 
the d$_{xy}$ orbital, an insulating state and also an antiferromagnetic (AFM) ordering 
to be stabilized, whereas a cubic FM metallic state is observed experimentally.  
This is the case also when LDA+U is applied for the structure relaxed with GGA.
For this reason, throughout the paper, we will use for structure 
optimizations of these multilayers including such a moderately correlated compound, 
the GGA-PBE functional.\cite{gga}
With the structure thus determined, we use
the LDA+U method for the calculations of the electronic and magnetic structure, and energy differences. 
In bulk SVO this procedure properly results in 
a FM half metal with 1 $\mu_B$/V atom.  


\section{General Considerations}

SrVO$_3$ has a lattice parameter of 3.84 \AA,\cite{srvo_struct} while the lattice parameter of SrTiO$_3$ 
(STO) is 3.90 \AA.\cite{stio} Since most of these superlattices are grown on a SrTiO$_3$ 
substrate, for closest comparison with experimental data we fix the $a$ lattice parameter 
to be the STO lattice parameter. However, we use the GGA-PBE exchange-correlation potential 
to optimize the superlattice $c$ parameter and also the internal coordinates of all the 
atoms by a force minimization together with a total energy minimization. 

Since the IF between SVO and STO has no polar discontinuity, the distortions introduced
at the interface between the two oxides by the nanostructuring can be understood as
first, strain due to the lattice parameter mismatch (1.5\%), and secondly, to charge
imbalance within the V $t_{2g}$ orbitals.  However, no ionic charge compensation effects
of the sort that are so interesting in LaAlO$_3$/SrTiO$_3$ nanostructures are
present.\cite{hwang,nakagawa,pentcheva2008,willmott,siemons,freeman,pentcheva2006}

We compare results for $m/n$ multilayers with $m$= 4 layers of STO 
(about 1.6 nm thickness) sandwiching an SVO layer with variable thickness from 1 to 5 
layers of SVO (0.4 to 2.0 nm), because we find 4 layers of STO sufficient to isolate
the SVO slabs to give two-dimensional behavior (negligible k$_z$-dispersion in the band structure). We analyze the evolution of 
the electronic structure for increasing thickness of SVO slabs ($n$ from 1 to 5).

\subsection{d$^1$ V ion in an octahedral environment}

\begin{figure}[ht]
\begin{center}
\includegraphics[width=\columnwidth,draft=false]{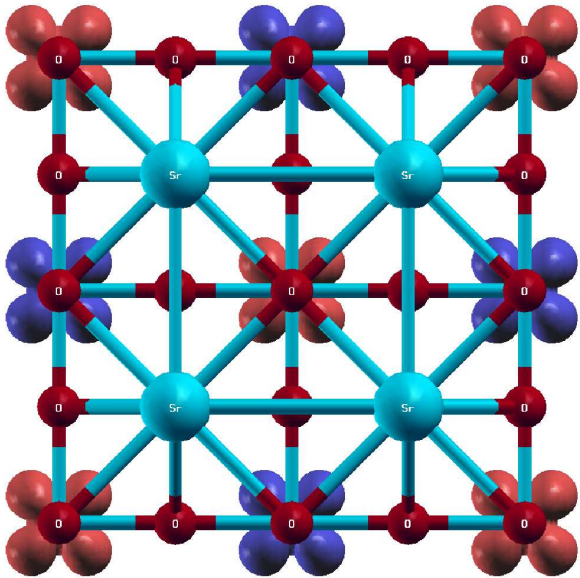}
\caption{(Color online) Spin density isosurface of the n=2 system at 0.8 e/\AA$^3$. 
Shown is a top view of the $x-y$ plane in the planar AF Mott insulator state, which
is the most stable only at larger values of U. The d$_{xy}$ orbital is occupied on
each V ion. Different colors indicate different spin directions.}\label{rho_4_2}
\end{center}
\end{figure}

One common feature within the SVO sublayer for all thicknesses we have studied is that
the V octahedral environment, especially at the interface (IF), 
will be tetrahedrally distorted from its
cubic symmetry in the bulk. It was noted above that such a distortion, 
treated fully (i.e. including structural relaxation) within the 
LDA+U method for reasonable values of U, produces an (incorrect) AF insulator state
for bulk SVO. In the case of the multilayers, our relaxation of the $c$ lattice parameter and atomic positions will lead 
to values of the interplane V-V distance somewhat smaller than the in-plane value.
The simplest scenario would be that the multilayer structuring of SVO on
an STO substrate will produce a tetragonal distortion of the oxygen octahedra around 
the V cations. This $c$-axis contraction of the octahedra leads to a preferred occupation 
of the d$_{xy}$ orbital. 
If $d_{xy}$ orbitals are occupied on all atoms in a layer, a small enough bandwidth or
strong enough intra-atomic interaction U will give AF ordering by superexchange. This 
effect is observed experimentally in similar multilayer systems with the d$^1$ compound 
being a Mott insulator like LaTiO$_3$.\cite{latio_ml}

A more realistic picture must allow for the likelihood that the degenerate
$d_{xz}, d_{yz}$ orbitals will have a narrower bandwidth in the $x-y$ plane, and correlation
effects may favor occupation of some combination of these orbitals. Even though small,
spin-orbit coupling breaks this degeneracy and specifies a favored combination. 
The actual occupation will depend on several factors. First of all is the tetragonal 
crystal field splitting of the $t_{2g}$ orbitals. 
In all cases we study, the $xy$ on-site energy is lower due to in-plane stretching
imposed by the STO substrate.  Second, the $x-y$ plane bandwidths of $d_{xy}$ and 
$d_{xz}, d_{yz}$ bands are
very different, and will depend on orbital order.
Thirdly, the orbital order
is closely tied to the magnetic configuration of the system. The simplest possibility (above), with 
all the electrons in an $d_{xy}$ orbital, leads to AF order (this orbital pattern can 
be seen in Fig. \ref{rho_4_2}), whereas we find that occupying $d_{xz}, d_{yz}$ orbitals 
favors FM order.

\subsection{Structural relaxation: strain effects}
The structural distortion as $n$ increases can be described in terms of lattice 
strain along the $c$-axis. Relaxing the $c$-axis value (and also the internal atomic 
positions) yields the results provided in Table \ref{tab_dist}. For understanding  
the structural distortions, we can define four different distances along the 
$c$-axis (within the plane they are constrained by the STO lattice parameter 3.905 \AA):
the V-V distance, the V-Ti distance across the IF, the Ti-Ti distance, and finally 
the average $c$ lattice parameter. We find that the 
Ti-Ti distance hardly changes, and also the V-Ti distance variation is minor. 
However, the V-V distance grows towards a limiting value as the number of SVO layers 
increases. Both V-V and Ti-Ti interlayer distances presented in the table vary only by 
$\pm$ 0.01 \AA\ for the various layers within each n/m system.

\begin{table}[h!]
\caption{Interlayer distances (\AA) between metal cations, for the various $n/m$ configurations under study. 
The important distances to describe the structure are Ti-Ti, Ti-V (across the interface), V-V, 
and the average c parameter.  The Ti-Ti separations, and V-V separations, for each
heterostructure are uniform to within 0.01 \AA.}\label{tab_dist}
\begin{center}
\begin{tabular}{|c|c|c|c|c|}
\hline
& Ti-Ti & V-Ti & V-V & c$_{av}$ \\
\hline
\hline

4/1 & 3.97 & 3.92 & --  & 3.95 \\
4/2 & 3.95 & 3.90 & 3.79 & 3.91 \\
4/3 & 3.95 & 3.91 & 3.83 & 3.91 \\
4/4 & 3.98 & 3.94 & 3.86 & 3.92 \\
4/5 & 3.98 & 3.93 & 3.87 & 3.92 \\
\hline
\end{tabular}
\end{center} 
\label{table1}
\end{table}

\subsection{Band lineups; intralayer supercells}
Some general features should be established. 
Due to the alignment of the SrVO$_3$ Fermi level (or gap) within the
SrTiO$_3$ gap between filled O $2p$ states and empty Ti $3d$ states, 
there is a 2.5 eV energy window which contains only the 
V $d$ bands that are of interest for us.
The oxygen $2p$ bands lie below the V $d$ bands.   
We note especially that we have
used laterally enlarged, $c(2\times 2)$ superstructures to allow the possibility
(or likelihood) of AF spin alignment as well as FM, and various orbital ordering patterns as well.
The ground states we obtain and analyze are all lower in energy than any that would have been
found in primitive $p(1\times 1)$ cells.

\subsection{Spin-orbit coupling}\label{soc}
SOC in the V atom usually produces minor effects, except for those which depend
entirely on it (such as magnetocrystalline anisotropy).  We find here however,
irrespective of the number of VO$_2$ layers, that SOC in conjunction with correlation
effects (the on-site repulsion U) completely alter the ground state that we obtain.

It has long been known that the $t_{2g}$ subshell provides a representation for 
$\ell$=1 (not $\ell$=2) orbital moments.\cite{epr,stevens,enough,lacroix,eschrig} 
The $m_{\ell}$ states are $d_0$ = $d_{xy}$, and $d_{\pm} =
d_{xz} \pm i d_{yz}$.  SOC splits these, with $d_{-}$ ($j_z=-\frac{1}{2}$)
lying lower.  This viewpoint is not common, because
the orbital moments are typically quenched by mixing with neighboring
orbitals.  Fairly recently several
cases have come to light in which $t_{2g}$ orbital moments can
be quenched\cite{khaliulin} by
structure-induced or spontaneous symmetry lowering, or even in a $5d$ system an orbital moment can
compensate a spin moment and prevent orbital ordering.\cite{kwanwoo} The calculations
presented here provide an additional example of the importance of SOC in $3d$
systems.

The strain in the
slabs we study breaks the $t_{2g}$ symmetry, with $d_{xy}$ lying lower.  
Strong magnetocrystalline anisotropy favors the magnetization along
the $c$-axis (by 130 meV/V compared to an in-plane orientation), making this the
quantization axis.  Then one can
expect competition between occupation of the $d_{-}$ and $d_{xy}$ orbitals in a $d^1$
ion, with kinetic energy (bandwidth and band placement) being a determining factor.
The energy gain, and thus
the large magnetocrystalline anisotropy, is related (see following sections)
to the formation of a large orbital
moment 0.75 $\mu_B$
along the $z$-axis when the $d_{-}$ orbital is occupied.  Since the spin moment is 
reduced somewhat from its atomic value of
1 $\mu_B$ by hybridization, the net magnetic moment on such an ion can be quite small.  

In the next section we show that an alternating orbitally-ordered (AOO) state, with half 
the electrons in a $d_{xy}$ orbital
and the other sublattice in a $d_{\pm}$ orbital, arises and leads to FM spin alignment. 
The AF solution becomes favored only at unphysically large values of U (above 5-6 eV).
In the reasonable range of values of U, this AOO state is
energetically favored for several SVO slab thicknesses. This AOO, Mott insulating FM 
state competes and overcomes the
Mott-insulating AF state, which has all the electrons in $xy$ orbitals. 
Unexpectedly, when SOC is included, the alternating orbital ordering produces an
FM Mott insulating state at small SrVO$_3$ thicknesses, below 2 nm, and
for realistic values of U for this multilayered system.  It is instructive to follow
the behavior through the insulator-to-metal transition with SVO thickness.

\section{Evolution of the electronic structure}

\begin{figure*}[ht]
\begin{center}
\includegraphics[width=0.98\columnwidth,draft=false]{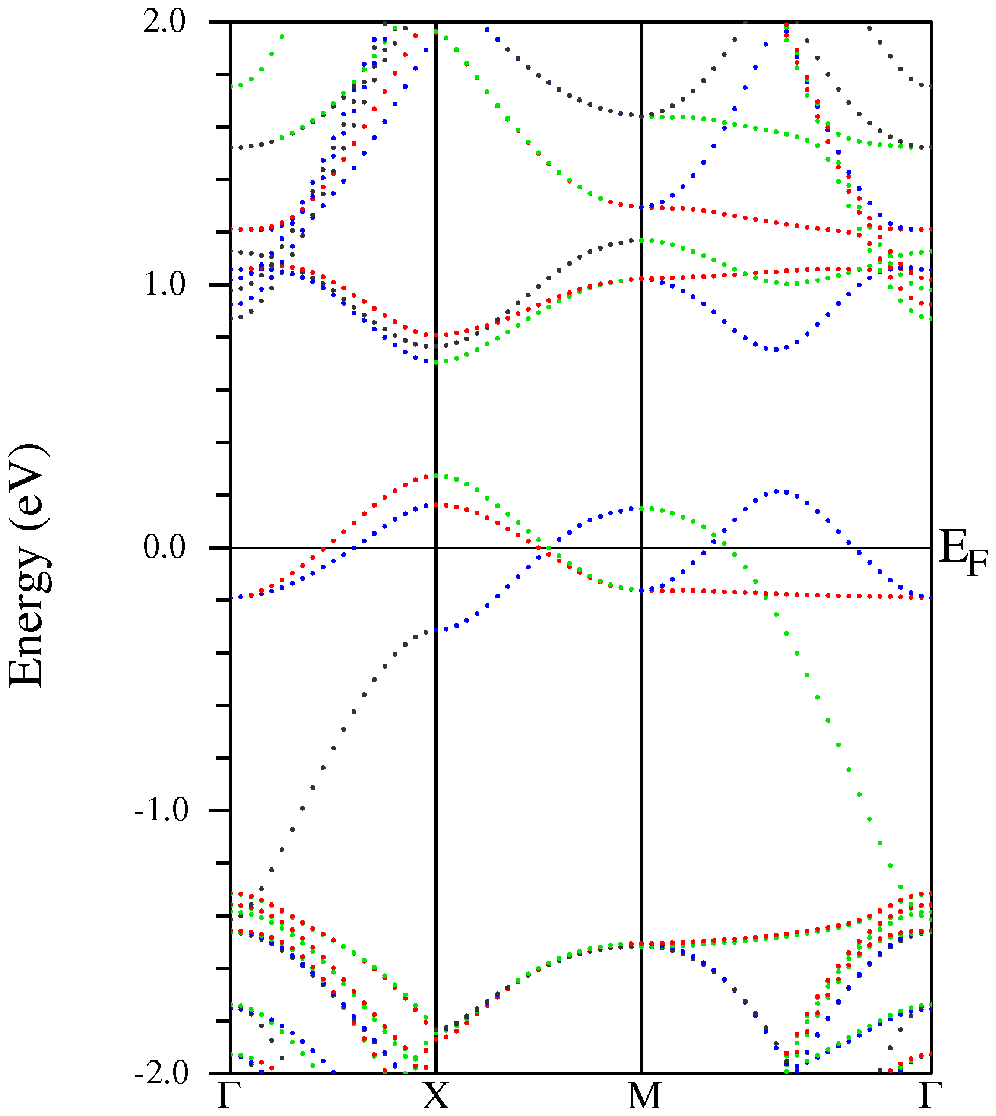}
\includegraphics[width=0.98\columnwidth,draft=false]{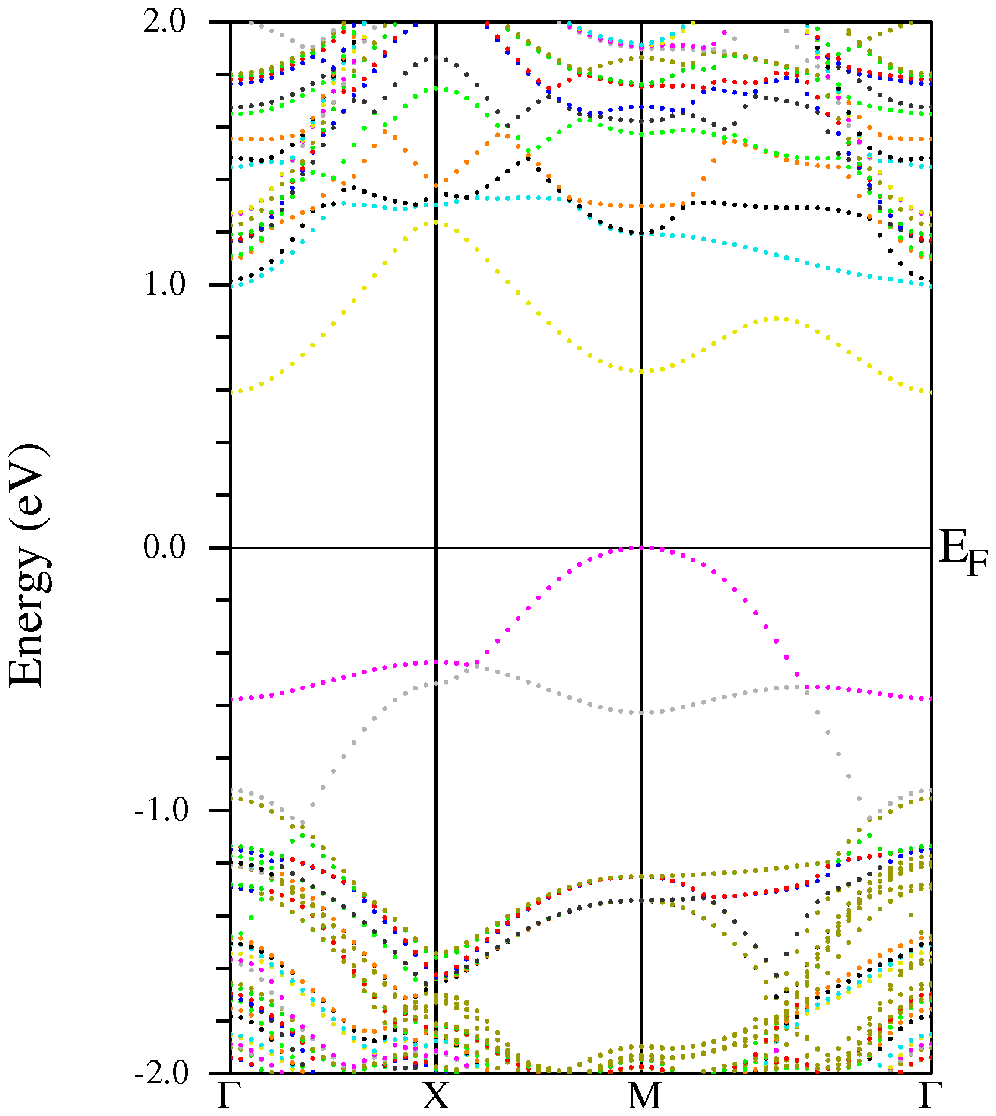}
\caption{(Color online) Majority spin LDA+U band structure of the n=1 system 
without (left panel) and with  
spin-orbit coupling (right panel) included. A SOC-driven FM Mott insulator regime 
occurs at this very small 
SVO thickness: the introduction of SOC breaks the $d_{xz}, d_{yz}$ symmetry,
and LDA+U splits the $d_{-}$ (down) and $d_{+}$ states (up),  
opening the gap. The lower and upper Hubbard bands lie at -0.5 eV and 1.3 eV, 
respectively.}\label{bs_4_1}
\end{center}
\end{figure*}

\subsection{n= 1 confined layer of SrVO$_3$}
When a single layer of SVO is confined by insulating STO, the resulting strain lowers the
$d_{xy}$ orbital energy, and it should be expected that a candidate ground state is
$d_{xy}$ ``orbitally ordered'' AF insulator due to superexchange.  Indeed this N\'eel
state can be obtained in our calculations, for the moderate range of U that is relevant.
The coupling arises through $dd\pi$-type hopping between $d_{xy}$ orbitals in the plane 
(see Fig. \ref{rho_4_2}) as expected from 
Goodenough-Kanamori-Anderson (GKA) rules.\cite{goody} However, the AOO phase with FM spin
alignment was also obtained, and it is lower in energy by 4 meV/V.  The two sublattices lead
to distinct sets of bands, as easily seen from the left panel of Fig. \ref{bs_4_1}.  
The $d_{xy}$ band has the familiar square-lattice shape (distorted
somewhat by 2nd neighbor coupling, and is 1.5 eV wide.  The $d_{xz} - d_{yz}$ bands are
much narrower (0.25 eV) because there is only $dd\pi$ coupling.  The centroid of the latter
pair lies about 0.5 eV above that of the $d_{xy}$ band, providing the magnitude of the splitting
of the $t_{2g}$ by strain.  The system is metallic, with all three bands leading to Fermi
surfaces. 

The picture is completely changed by SOC.
Figure \ref{bs_4_1} compares the majority spin band structure of the FM AOO state, first
without SOC, then with SOC included.  SOC has no effect on the $d_{xy}$ ($\ell_z = 0$)
band.  However, SOC breaks the symmetry of the $\ell_z = \pm 1$ doublet, and the narrow 
bandwidth (0.25 eV) compared to the value of U results in a Mott insulating type of 
splitting of the $d_{-}$ and $d_{+}$ bands, by roughly $\pm$U.  The SOC-driven symmetry
lowering is leveraged by the strong on-site interaction.  The result is an AOO FM Mott
insulator with a gap of 0.6 eV.
This  $d_{-}$ orbital acquires a large orbital moment of about 0.75 $\mu_B$, strongly compensating the
spin moment.  
This motif of FM AOO V orbitals will recur for thicker SVO slabs.

\begin{figure}[ht]
\begin{center}
\includegraphics[width=\columnwidth,draft=false]{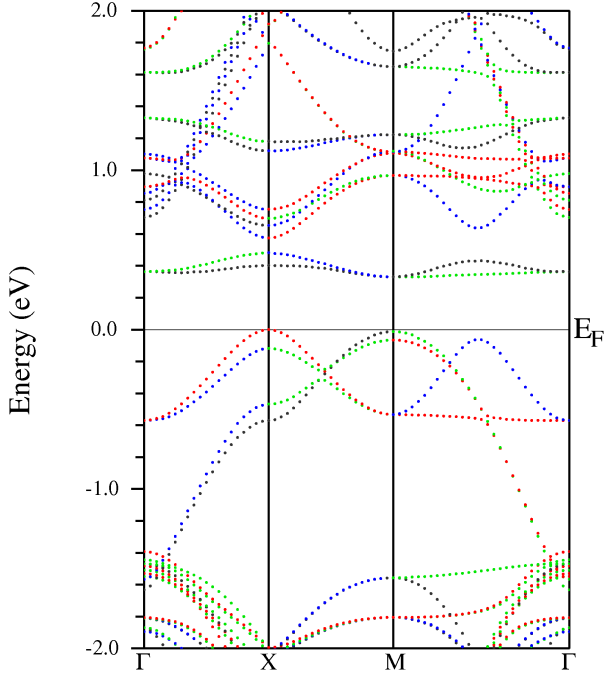}
\caption{(Color online) LDA+U band structure (without SOC) of the 4/2 system (U = 4.5 eV). 
A Mott-insulator regime occurs 
at this thickness of two SVO monolayers.} 
\label{bs_4_2}
\end{center}
\end{figure}

\subsection{n= 2 SrVO$_3$ layers}
The bands near the gap for  the two SVO layer slab are displayed in Fig. 
\ref{bs_4_2} without SOC included, to illustrate that for two layers there is already
a band gap without SOC, produced by interlayer coupling of $d_{xz}, d_{yz}$ states.
However, SOC produces the same orbital ordering and intralayer FM alignment as for n=1, and 
the layers are also spin-aligned to give an overall FM AOO Mott insulating
state. Comparing the total energy for
the $d_{xy}$  AF state with the more stable FM state, the energy difference is found to
be large, 76 meV/V. This energy difference includes the in-plane energy gain as for n=1
and the larger interplane stabilization due to a large dd$\sigma$ coupling between d$_-$ 
orbitals.

\subsection{n= 3 SrVO$_3$ layers}

\begin{figure}[ht]
\begin{center}
\includegraphics[width=\columnwidth,draft=false]{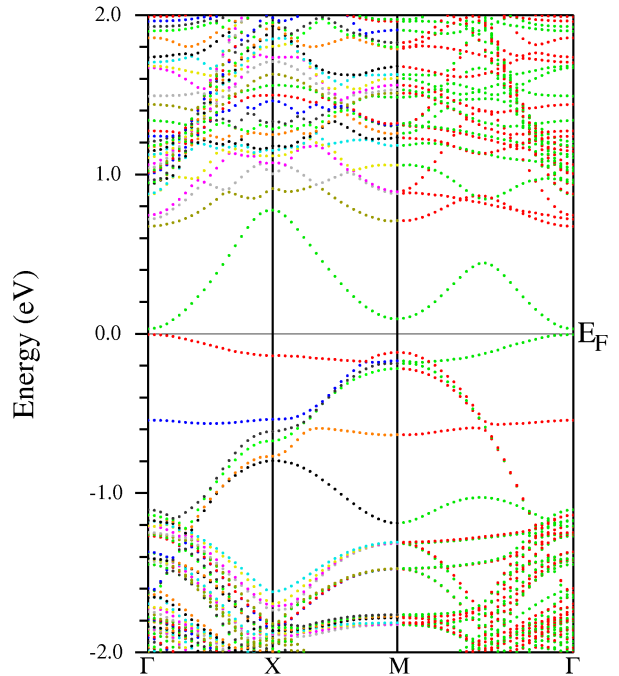}
\caption{(Color online) Band structure of the 4/3 system for U= 4.5 eV with spin-orbit coupling spin state. The minority spin state would be equivalent  in this AFM ground state. A very small gap insulator is obtained, on the verge of a metal-insulator transition. A metal is actually obtained at small U (below 4 eV).}\label{bs_4_3}
\end{center}
\end{figure}

As mentioned earlier, structural relaxation was always carried out using only the GGA
exchange-correlation functional.
We note that for this n=3 case GGA calculations give an AFM interplane coupling (up/down/up) of FM layers.
We carried out the structural 
relaxations in this magnetic structure; however, the type of magnetic order is not expected 
to affect the relaxation appreciably.

Returning to the LDA+U calculations to evaluate spin and orbital order, 
the ground state is FM overall (up/up/up)
with AOO order within the plane, and a  metallic band structure. 
As for n=1 and n=2, in-plane AFM ordering with all $d_{xy}$ orbitals
occupied can be obtained, but is energetically higher than the FM AOO state. In this case, 
we can compare the total energies of two configurations with the same in-plane orbital 
ordering, but different interlayer AOO alignment, AAA (like orbitals aligned along the
$c$-axis) or ABA. The AAA configuration 
is more stable by 6 meV/V, giving an idea of the strength of the interlayer coupling.

SOC again results in the same intralayer AOO FM state, however
in this case there is an extremely small gap at $\Gamma$, visible in Fig. \ref{bs_4_3}.
It differs from the n=2 system (and is similar to the n=1 system)
in having a gap only when SOC in included (breaking the
$d_{\pm}$ degeneracy that otherwise leaves half (or partially) filled bands).

\subsection{n=4 SrVO$_3$ layers}

\begin{figure}[ht]
\begin{center}
\includegraphics[width=\columnwidth,draft=false]{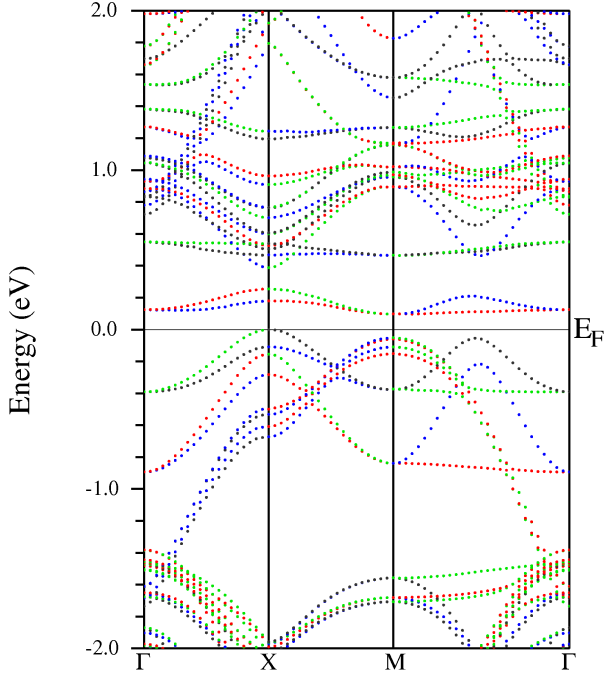}
\caption{(Color online) Majority spin band structure of the n=4 system (U= 4.5 eV) of 
this FM half-metallic, orbitally ordered ground state. The four characteristic, nearly
degenerate $d_{xy}$ bands lie in the -1.6 to -0.2 eV range. There is ferro-orbital ordering 
between planes.  The inter-layer coupling splits the $d_{-}$ bands by 0.2-0.3 eV.}\label{bs_4_4}
\end{center}
\end{figure}

The band structure of the n=4 system, which has the same AOO FM Mott insulating ground state, 
is shown neglecting SOC in 
Fig. \ref{bs_4_4} to show that SOC is not necessary to produce a gap (as for n=2), although it
does change the band structure.  The four nearly degenerate $d_{xy}$ bands with their simple square-lattice
shape lie in the -1.6 to -0.1 eV range, interlayer coupling is very small for these $d_{xy}$
orbitals.
The four band pairs for the $d_{xz}, d_{yz}$ orbitals correspond to linear combinations of 
similarly shaped bands for  each of the four layers.  The band dispersion decreases from
0.7 eV for the lowest band to zero for the highest (unoccupied) bands. 
By using a simple tight-binding model with interplane coupling, the band ordering and hence the
occupation of the bands is readily reproduced.

\begin{figure}[ht]
\begin{center}
\includegraphics[width=0.45\columnwidth,draft=false]{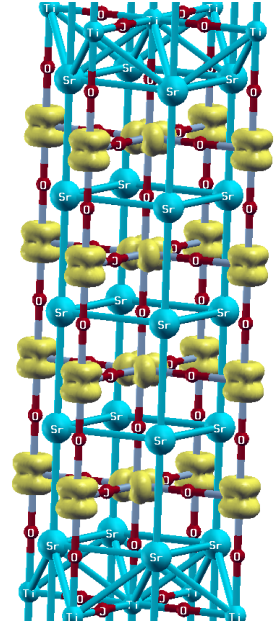}
\caption{(Color online) Spin density isosurface of the n=4 system (U=4.5 eV) of the 
majority spin electrons of this FM insulating solution. The in-plane alternating orbital 
ordering $d_{xy}$ and $d_{-}$ orbitals is apparent, and also the interlayer orbital 
configuration.  The stabilization of the ferro-orbital occupation along the $z$-axis
is stabilized by the interaction  between V d$_{xz \pm iyz}$ orbitals in this FM ground 
state.}\label{rho_4_4}
\end{center}
\end{figure}

\begin{figure}[ht]
\begin{center}
\includegraphics[width=\columnwidth,draft=false]{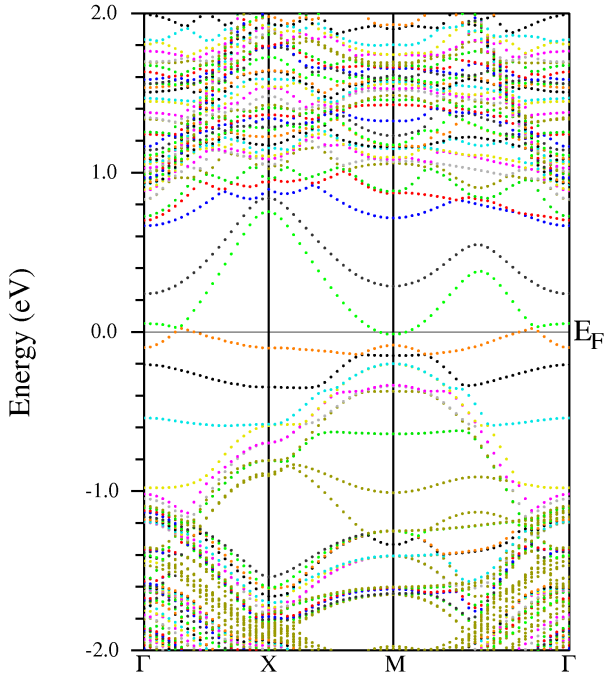}
\caption{(Color online) Band structure of the 4/5 system (U=4.5 eV) of the majority spin 
of this FM half-metallic state.  
There is orbital ordering in-plane, and ferro-orbital ordering
between layers, as for thinner SrVO$_3$ slabs. However, with the added bands, the interlayer coupling is no longer large
enough to open a gap, leaving a semimetallic state with band overlap at $\Gamma$.}\label{bs_4_5}
\end{center}
\end{figure}

\subsection{n=5 SrVO$_3$ layers}
For n=5 several orbital orderings can be obtained and their energies compared.  Of the
solutions we studied, again the AOO FM layers, spin aligned between layers (globally
FM) is again the ground state.  Tetragonal strain and SOC result in the same AOO arrangement
and corresponding identifiable bands, which are shown in Fig. \ref{bs_4_5}.  The five $d_{xy}$
bands have a total splitting no more than 0.4 eV, and are completely filled.  The five $d_{-}$ 
bands, each with small bandwidth, cover a range of 1.4 eV.  The lower two (-1.4 eV to -0.9 eV)
mix with the O $2p$ bands, and are not as obvious as the upper three bands.  

The distinction for n=5 is that the uppermost $d_{-}$ band is not fully occupied, but overlaps
by 0.2 eV a conduction band that dips below the $d_{-}$ at the $\Gamma$ point and leads to
a semimetal. The insulator-to-metal transition has occurred between n=4 and n=5, very similar
to the transition observed by Kim {\it et al.}\cite{svo_sto_ssc} in their SVO/STO multilayers.
Unlike the VO$_2$/TiO$_2$ system where the transition proceeds through a point Fermi surface,
semi-Dirac phase,\cite{sD_prl,swapno_sD,vo2_tio2_mit} this transition appears to occur in 
the classic fashion of band overlap.
The strong effects of tetragonal strain (symmetry breaking) and SOC (further symmetry breaking)
are finally overcome by the increasing delocalization across the SVO slab as the quantum
confinement effects are eroded.

\section{Transition through a singular Fermi surface}
\begin{figure}[ht]
\begin{center}
\includegraphics[width=\columnwidth,draft=false]{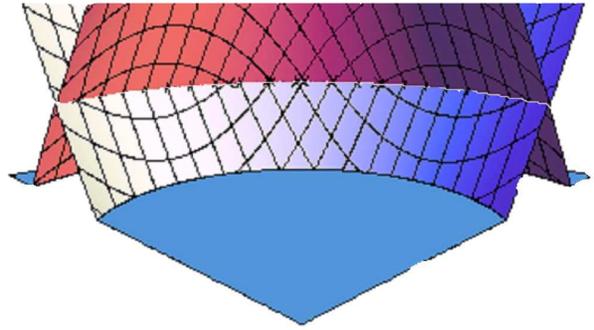}
\caption{(Color online) Two (locally) linearly dispersive bands in the form $\epsilon_k$= $\pm$ 
v$\left|\left|k\right|-k_F\right|$ come together forming a continuous circle of 
Dirac points, {\it i.e.} a ``Dirac circle.'' Only one quadrant of the zone is displayed.
This model represents the two bands 
closest to the Fermi level for the n= 5 system displayed in Fig. \ref{bs_4_5}.}\label{dirac_circles}
\end{center}
\end{figure}

A very unconventional insulator-to-metal transition nearly occurs at the n=5 thickness, and
might actually occur for somewhat different value of intra-atomic repulsion U or different
value of strain.  If some small change lifted the conduction band at the M point in Fig.
\ref{bs_4_5} above the Fermi level (it overlaps only slightly as it is), and if the 
overlapping valence and conduction bands at
$\Gamma$ are in the (small $k$) quadratic limit so their constant energy surfaces are circles,
then the equality of electron and hole density leads to {\it coinciding} electron and hole 
Fermi surfaces.  The ``Fermi surface'' is actually two identical electron and hole Fermi lines,
the ``Fermi
surface'' has a boundary but no area.  
The bands near the Fermi energy, given in the simplest model by
\begin{eqnarray}
\epsilon_k = \pm v (|\vec k| - k_F),
\end{eqnarray}
are presented in Fig. \ref{dirac_circles}, where these linearly dispersing bands are shown.  With doping,
the electron and hole surfaces (lines, since this is two dimensions) separate, leaving an annulus
that contains electrons if electron-doped, or holes if it is hole-doped.  It is thus easy to
understand how the annulus vanished as the doping level vanishes. The situation is in fact a
continuum of radial Dirac points.  Conversely, it presents the limit of a semi-Dirac 
point\cite{sD_prl,swapno_sD} when the
effective mass diverges.

From Fig. \ref{bs_4_5}, it is clear that (with the conduction band at M out of the picture)
the positioning of the Fermi level at the band
crossing {\it point} is topologically determined: only for precisely that Fermi level are
an integral number of bands occupied, which is exactly what is required to occupy the V
$3d$ electrons.  The lowering of the symmetry of the eigenstates off of the symmetry 
directions (corresponding to some subgroup of the group of the symmetric k-point) may
lead to coupling of the bands and the opening of a gap away from the X or M points (or both).
In this case, one is left with four (respectively, eight) semi-Dirac points along symmetry
lines.  

However, it may occur that the bands have different symmetry throughout the Brillouin zone,
in which case no gap opens.  The simplest example is even and odd symmetry under $z$-reflection,
$d_{xy}$ states for example being even, and $d_{xz}, d_{yz}$ states being odd.  This is the
case in which this Dirac-continuum of points may arise.

\section{Summary}

We have analyzed the transition from insulator to metal in the electronic structure of multilayers 
(SrTiO$_3$)$_4$/(SrVO$_3$)$_n$, with $n$ varying from 1 to 5.  The transition is observed to occur
between n=4 and n=5. The origin of the observed changes is surprisingly intricate, with tetragonal
strain and spin-orbit coupling (each with an associated symmetry breaking) leveraging strong
interaction effects modeled by the LDA+U approach.  The effects of quantum confinement finally 
determine the conduction character.   Insulating behavior with a peculiar alternating orbital ordering 
within each V layer and FM magnetic order results from a ferromagnetic Mott insulating state for
n=4 or less.   Ferromagnetic Mott insulators are rare, and these results indicate how this kind
may be achieved (even designed) in oxide nanostructures.  
The FM insulator-to-FM metal transition finally results from band overlap as quantum confinement
effects decrease.  This system is very close to, if not at, an unusual semimetal state for n=5
in which the Fermi surface is topologically determined and consists of two degenerate electron
and hole circles around the k=0 point.

\section{Acknowledgments}

This project was supported by DOE grant DE-FG02-04ER46111 and through interactions with
the Predictive Capability for Strongly Correlated Systems team of the Computational
Materials Science Network.


\end{document}